\documentclass[prl,twocolumn,amsmath,amssymb]{revtex4}
\usepackage{graphicx}
\usepackage{subfigure}

\begin{document}

\title{Supplementary Methods for \\
``Comment on `Nuclear Emissions During Self-Nucleated Acoustic Cavitation' ''}
\author{B. Naranjo}
\affiliation{UCLA Department of Physics and Astronomy,
Los Angeles, California 90095, USA} 
\maketitle

\newcommand{\chilambda}{\ensuremath{\chi^2_{\lambda,p}}}
\newcommand{\isotope}[2]{\mbox{$^{#2}\mbox{#1}$}}

\paragraph{Monte Carlo methods.}

\begin{figure*}
  \centering
    \includegraphics[width=100mm]{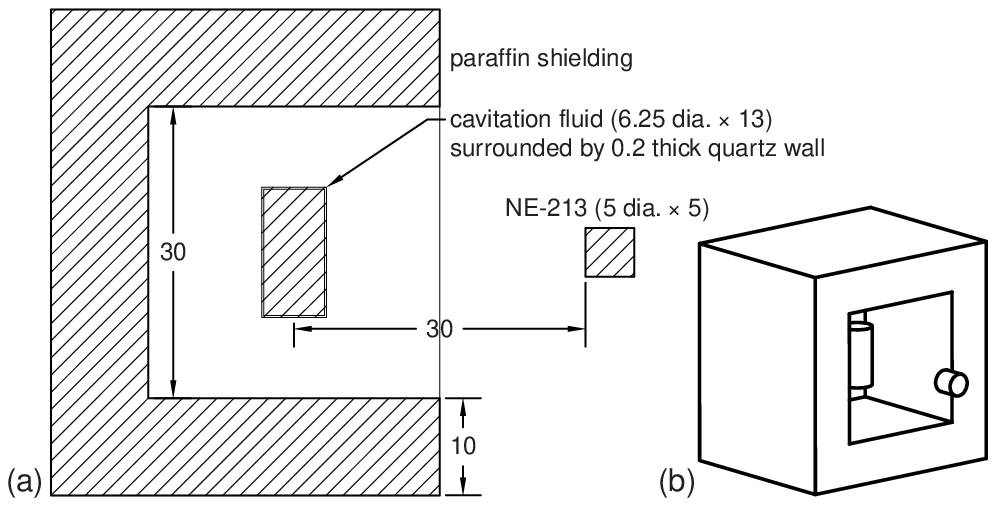}
    \caption{Monte Carlo geometry.  All dimensions in cm.
(a) Section view. (b) Perspective view.}
    \label{fig:geometry}
\end{figure*}

Figure~\ref{fig:geometry} shows the geometry used in the ``2.45 MeV
w/shielding'' detector response simulation.  The cavitation fluid,
modeled as described in footnote 13 of
\cite{taleyarkhan2006supp}, consists of carbon, deuterium, chlorine,
oxygen, nitrogen, and uranium. The simulation, using
\textsc{Geant4} \cite{agostinelli2003}, includes all relevant
neutron interactions, particularly elastic scattering, thermal
capture, $(n, n' \gamma)$ scattering, and neutron-induced fission.

To calculate the response function, neutrons of energy 2.45 MeV,
emitted isotropically from the center of the flask, scatter through
the materials.  When a neutron elastically scatters protons in the
liquid scintillator, the recoil energies are converted to equivalent
electron energies \cite{verbinski1968}, summed, and then smeared
according to the detector's resolution function, eventually obtaining
the response function $g(E_{ee})$.

In the same manner, I calculate the other response functions with the
following changes: the ``2.45 MeV'' simulation does not include the
paraffin shield, and the radioisotope \cite{lajtai1990,anderson1972}
simulations ``Cf-252'' and ``PuBe'' assume there are no intervening
scattering materials between the sources and the detector.

\paragraph{Statistical methods.}

Following the notation of \cite{baker1984}, the raw data from Fig.~9(b)
of \cite{taleyarkhan2006supp} are
\begin{eqnarray*}
  n_k &=& \text{``cavitation off'' counts in channel } k \\
  n'_k &=& \text{``cavitation on'' counts in channel } k. \\
\end{eqnarray*}
Each run is 300 s in duration, and Fig.~4 of \cite{taleyarkhan2006}
shows the background-subtracted signal $n'_k - n_k$.

The $n_k$ background data are modeled by a sum of two exponentials,
and the $n'_k$ data are modeled by the same background function plus
the scaled response function,
\begin{eqnarray*}
  y_k &=&  A_1 \exp(-k/A_2) + A_3\exp(-k/A_4) \\
  y'_k &=& y_k + A_5 \, g_k. \\
\end{eqnarray*}
The binned response function $g_k$, is found by averaging $g(E_{ee})$
over the energy range of channel $k$.

\begin{figure*}
  \centering
  \subfigure{
    \includegraphics[width=83mm]{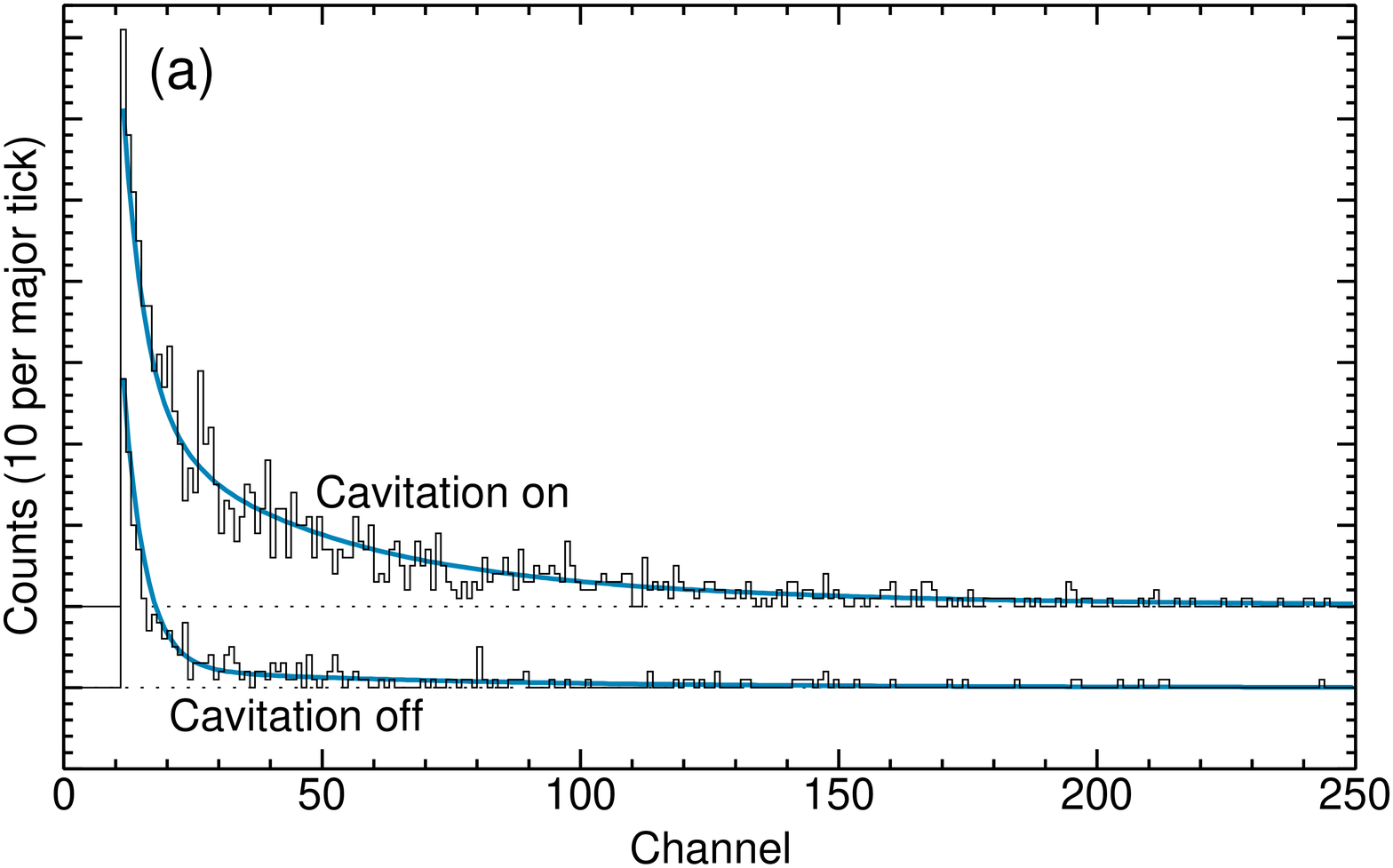}}
  \subfigure{
    \includegraphics[width=83mm]{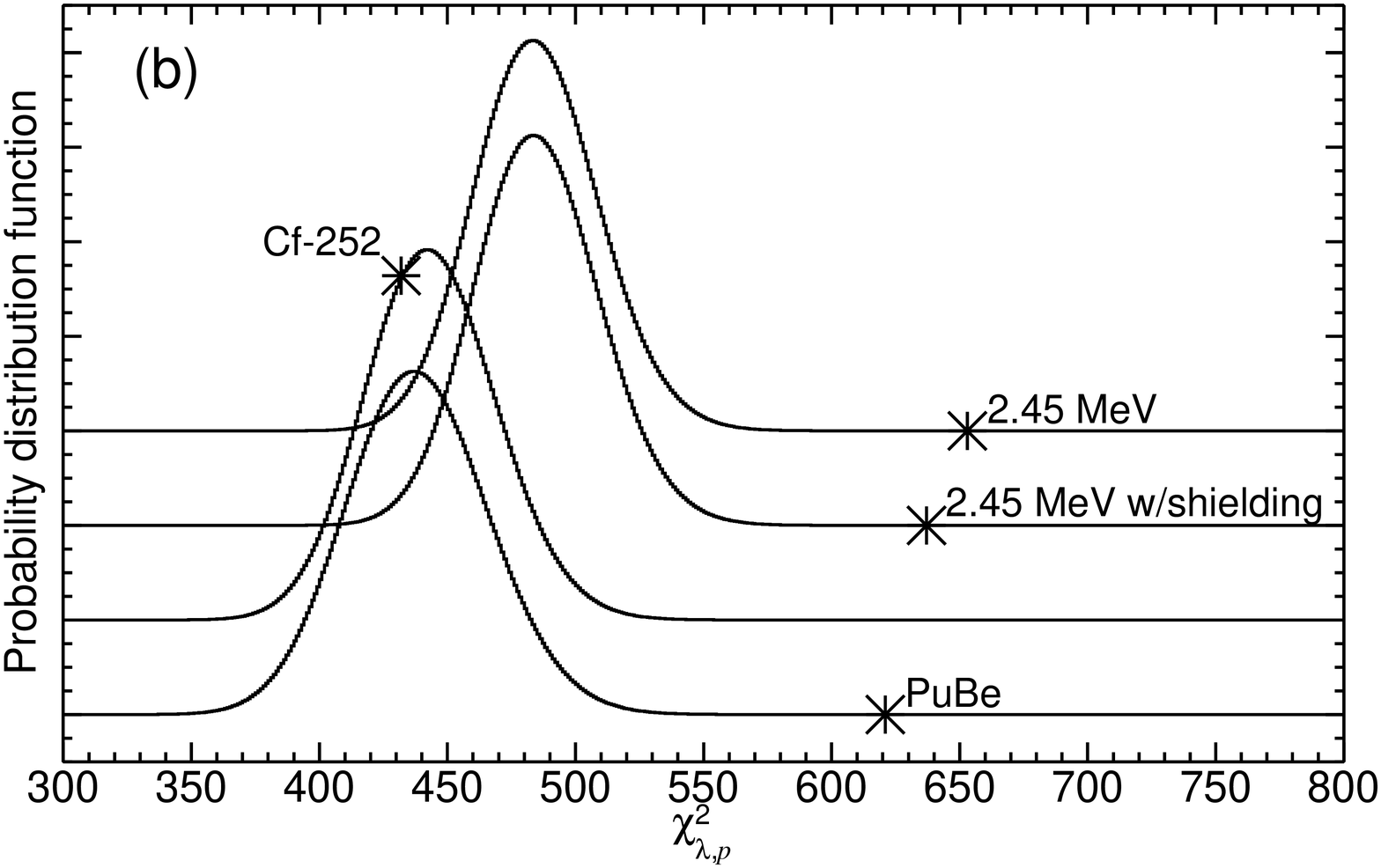}}
  \caption{Fit statistics.
(a) Fit using the simulated \isotope{Cf}{252} response function for
$g_k$.  The histograms are $n_k$ and $n'_k$, and the smooth blue lines
are theoretical curves $y_k$ and $y'_k$.  The minimized value of
\chilambda~is 432.  For clarity, the two graphs are offset by ten
counts.
(b) Numerically sampled distributions of \chilambda~for the four
hypotheses.  Observed values of \chilambda~are shown. }
    \label{fig:statistics}
\end{figure*}

Then, the Poisson likelihood chi-square \cite{baker1984} is
\[
  \chilambda = 2 \sum_{k=11}^{249} \left[ \phi_k + \phi'_k \right],
\]
where
\begin{eqnarray*}
  \phi_k &=& y_k - n_k + n_k \ln (n_k/y_k) \\
  \phi'_k &=& y'_k - n'_k + n'_k \ln (n'_k/y'_k). \\
\end{eqnarray*}
Note that, under proper conditions \cite{baker1984},
\chilambda~asymptotically approaches a $\chi^2$ distribution.  Moreover,
better fits give lower values of \chilambda.  Minimization
\cite{james1975} of \chilambda~determines the five fit parameters
$A_i$.  See Fig.~\ref{fig:statistics}(a) for the fit using the
``Cf-252'' response function.

To determine the distribution $f(\chilambda)$ for a given fit, I
sample from many synthetic data sets, each chosen, for $k=11, \ldots,
249$, from Poisson distributions of mean value $y_k$ and $y'_k$ .  In
the Comment, I report the goodness-of-fit as a Z-value, defined by
\[
   \int_{\text{obs.~} \chilambda}^{\infty} f(\chi^2) \, d\chi^2 =
   \frac{1}{\sqrt{2 \pi}} \int_{Z}^{\infty} e^{-t^2/2} \, dt,
\]
which expresses the observed value of \chilambda~in terms of the
equivalent number of standard deviations from the mean of a normal
distribution.  As shown in Fig.~\ref{fig:statistics}(b), the observed
value of \chilambda~for the ``Cf-252'' fit is within one equivalent
standard deviation and is therefore statistically consistent.  The
other three fits are outside five equivalent standard deviations, and
are therefore statistically inconsistent.


\begin{thebibliography}{8}
\expandafter\ifx\csname natexlab\endcsname\relax\def\natexlab#1{#1}\fi
\expandafter\ifx\csname bibnamefont\endcsname\relax
  \def\bibnamefont#1{#1}\fi
\expandafter\ifx\csname bibfnamefont\endcsname\relax
  \def\bibfnamefont#1{#1}\fi
\expandafter\ifx\csname citenamefont\endcsname\relax
  \def\citenamefont#1{#1}\fi
\expandafter\ifx\csname url\endcsname\relax
  \def\url#1{\texttt{#1}}\fi
\expandafter\ifx\csname urlprefix\endcsname\relax\def\urlprefix{URL }\fi
\providecommand{\bibinfo}[2]{#2}
\providecommand{\eprint}[2][]{\url{#2}}

\bibitem[{tal()}]{taleyarkhan2006supp}
\bibinfo{note}{R.~P.~Taleyarkhan {\it et al.}, EPAPS Document No.
  E-PRLTAO-96-019605}.

\bibitem[{\citenamefont{Agostinelli et~al.}(2003)\citenamefont{Agostinelli,
  Allison, Amako, Apostolakis, Araujo, Arce, Asai, Axen, Banerjee, Barrand
  et~al.}}]{agostinelli2003}
\bibinfo{author}{\bibfnamefont{S.}~\bibnamefont{Agostinelli}} {\it et al.},
  \bibinfo{journal}{Nucl. Instr. and Meth. A}
  \textbf{\bibinfo{volume}{506}}, \bibinfo{pages}{250} (\bibinfo{year}{2003}).

\bibitem[{\citenamefont{Verbinski et~al.}(1968)\citenamefont{Verbinski, Burrus,
  Love, Zobel, Hill, and Textor}}]{verbinski1968}
\bibinfo{author}{\bibfnamefont{V.~V.} \bibnamefont{Verbinski}},
  \bibinfo{author}{\bibfnamefont{W.~R.} \bibnamefont{Burrus}},
  \bibinfo{author}{\bibfnamefont{T.~A.} \bibnamefont{Love}},
  \bibinfo{author}{\bibfnamefont{W.}~\bibnamefont{Zobel}},
  \bibinfo{author}{\bibfnamefont{N.~W.} \bibnamefont{Hill}}, \bibnamefont{and}
  \bibinfo{author}{\bibfnamefont{R.}~\bibnamefont{Textor}},
  \bibinfo{journal}{Nucl. Instr. and Meth.} \textbf{\bibinfo{volume}{65}},
  \bibinfo{pages}{8} (\bibinfo{year}{1968}).

\bibitem[{\citenamefont{Lajtai et~al.}(1990)\citenamefont{Lajtai, Dyachenko,
  Kononov, and Seregina}}]{lajtai1990}
\bibinfo{author}{\bibfnamefont{A.}~\bibnamefont{Lajtai}},
  \bibinfo{author}{\bibfnamefont{P.~P.} \bibnamefont{Dyachenko}},
  \bibinfo{author}{\bibfnamefont{V.~N.} \bibnamefont{Kononov}},
  \bibnamefont{and} \bibinfo{author}{\bibfnamefont{E.~A.}
  \bibnamefont{Seregina}}, \bibinfo{journal}{Nucl. Instr. and Meth. A}
  \textbf{\bibinfo{volume}{293}}, \bibinfo{pages}{555} (\bibinfo{year}{1990}).

\bibitem[{\citenamefont{Anderson and Neff}(1972)}]{anderson1972}
\bibinfo{author}{\bibfnamefont{M.~E.} \bibnamefont{Anderson}} \bibnamefont{and}
  \bibinfo{author}{\bibfnamefont{R.~A.} \bibnamefont{Neff}},
  \bibinfo{journal}{Nucl. Instr. and Meth.} \textbf{\bibinfo{volume}{99}},
  \bibinfo{pages}{231} (\bibinfo{year}{1972}).

\bibitem[{\citenamefont{Baker and Cousins}(1984)}]{baker1984}
\bibinfo{author}{\bibfnamefont{S.}~\bibnamefont{Baker}} \bibnamefont{and}
  \bibinfo{author}{\bibfnamefont{R.~D.} \bibnamefont{Cousins}},
  \bibinfo{journal}{Nucl. Instr. and Meth.} \textbf{\bibinfo{volume}{221}},
  \bibinfo{pages}{437} (\bibinfo{year}{1984}).

\bibitem[{\citenamefont{Taleyarkhan et~al.}(2006)\citenamefont{Taleyarkhan,
  West, {Lahey,~Jr.}, Nigmatulin, Block, and Xu}}]{taleyarkhan2006}
\bibinfo{author}{\bibfnamefont{R.~P.} \bibnamefont{Taleyarkhan}},
  \bibinfo{author}{\bibfnamefont{C.~D.} \bibnamefont{West}},
  \bibinfo{author}{\bibfnamefont{R.~T.} \bibnamefont{{Lahey,~Jr.}}},
  \bibinfo{author}{\bibfnamefont{R.~I.} \bibnamefont{Nigmatulin}},
  \bibinfo{author}{\bibfnamefont{R.~C.} \bibnamefont{Block}}, \bibnamefont{and}
  \bibinfo{author}{\bibfnamefont{Y.}~\bibnamefont{Xu}}, \bibinfo{journal}{Phys.
  Rev. Lett.} \textbf{\bibinfo{volume}{96}}, \bibinfo{eid}{034301}
  (\bibinfo{year}{2006}).

\bibitem[{\citenamefont{{James} and {Roos}}(1975)}]{james1975}
\bibinfo{author}{\bibfnamefont{F.}~\bibnamefont{{James}}} \bibnamefont{and}
  \bibinfo{author}{\bibfnamefont{M.}~\bibnamefont{{Roos}}},
  \bibinfo{journal}{Comput. Phys. Comm.} \textbf{\bibinfo{volume}{10}},
  \bibinfo{pages}{343} (\bibinfo{year}{1975}).

\end{thebibliography}

\end{document}